# CT-based Subchondral Bone Microstructural Analysis in Knee Osteoarthritis via MR-Guided Distillation Learning

Yuqi Hu, BS[1], Xiangyu Zhao, BS[1], Gaowei Qing, MD, [2], Kai Xie, PhD[3], Chenglei Liu, MD[2]✉, Lichi Zhang, PhD[1]✉

**Author Affiliations**

[1]School of Biomedical Engineering, Shanghai Jiao Tong University, China.

[2]Department of Radiology, Shanghai Ninth People's Hospital, Shanghai Jiao Tong University School of Medicine, Shanghai, China.

[3]Shanghai Key Laboratory of Orthopaedic Implants, Department of Orthopaedic Surgery, Shanghai Ninth People's Hospital, Shanghai Jiao Tong University School of Medicine, Shanghai, China.

**The name and street address of the institution from which the work originated**

Name: Shanghai Jiao Tong University

Street Address: No. 1954 Huashan Road, Xuhui District, Shanghai, China

**Co-senior Authors**

Lichi Zhang: Institute for Medical Imaging Technology, School of Biomedical Engineering, Shanghai Jiao Tong University, China. Email: lichizhang@sjtu.edu.cn

Or Chenglei Liu: Department of Radiology, Shanghai Ninth People's Hospital, Shanghai Jiao Tong University School of Medicine, Shanghai, China.

Email: lcl1984@aliyun.com, phone number 15800585921. Zhizaoju road 639#, Huangpu district, Shanghai ,200011.

**Manuscript Type**

Original Research

**Word count for text**

2999

**Number of figures and tables**

5 figures and 4 tables

**Data sharing statement**

Data generated or analyzed during the study are available from the corresponding author by request.

# Abbrivated Title Page:

**Title of the manuscript:** CT-based Subchondral Bone Microstructural Analysis in Knee Osteoarthritis via MR-Guided Distillation Learning

**Article Type: Original Research**

**Summary:** The MR-guided distillation learning significantly enhances CNN-based trabecular parameter regressions on low-resolution CT, and demonstrates the feasiblity of CT-based subchondral bone microstructural analysis in knee osteoarthritis.

**Key Results**

(1) For subchondral bone analysis, the trabecular parameters regressed by our CT-based SRRD method have a strong correlation with MR-based ground-truth.

(2) By incorporating the distillation learning technique into the proposed SRRD method, we have shown a significant improvement in the performance of trabecular parameters regression and KOA classification based on CT images.

(3) Our SRRD method implements subchondral bone analysis using easily acquired multi-detector CT (MD-CT) images, which demonstrates great efficiency and cost-effectiveness.

# Abbreviations

KOA = knee osteoarthritis, TRE = target registration errors, ICC = intraclass correlation coefficients, BV/TV = bone volume/total volume, Tb.Th = trabecular thickness, Tb.Sp = trabecular separation, Tb.N = trabecular number, SRRD = synthesis, registration, regression and diagnosis, KL = Kellgren & Lawrence, AUC = area under the receiver operating characteristic curve.

# Abstract

**Background:** MR-based subchondral bone effectively predicts knee osteoarthritis. However, its clinical application is limited by the cost and time of MR.

**Purpose:** We aim to develop a novel distillation-learning-based method named SRRD for subchondral bone microstructural analysis using easily-acquired CT images, which leverages paired MR images to enhance the CT-based analysis model during training.

**Materials and Methods:** Knee joint images of both CT and MR modalities were collected from October 2020 to May 2021. Firstly, we developed a GAN-based generative model to transform MR images into CT images, which was used to establish the anatomical correspondence between the two modalities. Next, we obtained numerous patches of subchondral bone regions of MR images, together with their trabecular parameters (BV/TV, Tb.Th, Tb.Sp, Tb.N) from the corresponding CT image patches via regression. The distillation-learning technique was used to train the regression model and transfer MR structural information to the CT-based model. The regressed trabecular parameters were further used for knee osteoarthritis classification.

**Results:** A total of 80 participants (mean age, 51 years ± 15 [SD]; 35 female, 45 male) were evaluated. CT-based regression results of trabecular parameters achieved intra-class correlation coefficients (ICCs) of 0.804±0.037, 0.773±0.042, 0.711±0.063, and 0.622±0.133 for BV/TV, Tb.Th, Tb.Sp, and Tb.N, respectively. The use of distillation learning significantly improved the performance of the CT-based knee osteoarthritis classification method using the CNN approach, yielding an AUC score of 0.767 (95% CI, 0.681-0.853) instead of 0.658 (95% CI, 0.574-0.742) ($p<.001$).

**Conclusions:** The proposed SRRD method showed high reliability and validity in MR-CT registration, regression, and knee osteoarthritis classification, indicating the feasibility of subchondral bone microstructural analysis based on CT images.

# Main Body

## Introduction

Knee Osteoarthritis (KOA) is a prevalent degenerative joint disease that causes pain and disability in elder individuals (1, 2). Increasing evidence suggests that OA is a whole joint disease (3), and the significance of subchondral bone in the onset and progression of OA has been established (2). Besides, the progress of KOA and its associated symptoms are also closely linked with the changes in subchondral trabecular architecture, which can be traced through the radiographic features from related imaging biomarkers (4-6). Consequently, imaging-derived subchondral trabecular biomarkers have the potential to assist in the early diagnosis of KOA (7).

Accurately quantifying trabecular biomarkers in clinical settings can be challenging, especially considering the trabecular structures in human bones range from 50 to 200μm (8). Several studies have utilized high-resolution quantitative (HR-QCT) or micro-computed tomography (μCT) scans to investigate trabecular microstructure (9, 10). However, employing these techniques for quantitative assessment of trabecular architecture in the clinical scenario is impractical, due to the limited availability and the high radiation doses. High-resolution MRI is considered as the optimal modality with clear observations of cartilage and subchondral bone (11). MR-derived trabecular parameters correlated with histology (12), μCT (12), and biomechanical strength (13) derived from in vitro study. MR-based subchondral trabecular biomarkers also predicted KOA pain and radiographic progression (7, 14). However, MRI encounters barriers such as longer acquisition time, higher imaging cost, and more complex post-processing methods, which hinder the image collecting and clinical application of MR-based subchondral bone analysis (15).

Multi-Detector CT (MD-CT) provides highly detailed cross-sectional anatomical images and is considered standard in clinical use for the evaluation and management of KOA and arthroplasty (16). Given the limitations inherent in the above-mentioned high-resolution imaging techniques, accessing trabecular biomarkers from MD-CT scans provides an opportunity for evaluating these biomarker changes. However, with slice thickness on the order of 500μm and in-plane resolution of ~150μm, imaging of trabecular microstructure is certainly limited and subjected to substantial partial volume effects (16). Knowledge distillation(17), a highly useful technique in artificial intelligence, is capable of harness the features of high-resolution modalities to guide the analysis of low-resolution images(18). Building on its wide application in medical image segmentation (17, 18), we propose to utilize this technique between MR and CT to assist CT-based subchondral bone microstructural analysis.

In this paper, we hypothesize that distillation learning from high-resolution MR to CT can improve CT-based subchondral bone microstructural analysis. We aim to develop a novel CT-based subchondral bone microstructural analysis method named SSRD, and investigate its feasibility in obtaining trabecular parameters. Trabecular parameters calculated on MR images serve as ground-truth of trabecular biomarkers, and our main task is to regress them from the corresponding CT images. Our main contributions include two parts: (1) employing the distillation learning technique to enhance the trabecular-parameter-regression performance of convolutional neural networks (CNNs). It pretrains an MR-based teacher model to guide the CT-based student regression model through distillation loss functions, which passes knowledge learned on MR to CT and vastly improves the regression performance. (2) proposing a synthesis-based unsupervised registration module, which eliminates space gaps between MR and CT to access anatomical correspondent image patches. These patches serve as inputs for the distillation-learning-based regression module. The proposed synthesis and registration module performs MR-to-CT transformation without paired data and conducts coarse-to-refine registration in a mono-modal way. With the foundation of accurate registration to align modalities, distillation-learning becomes possible for the enhancement of CT-based models in the training time.

## Materials and Methods

### Patient Dataset

This study was approved by the institutional review board of XX(No.SH9H-2020-T395-2). Written informed consent was obtained from all participants. An orthopedic surgeon (XX) with fifteen years of experience recruited 149 consecutive participants from October 2020 to May 2021. Of these, a subset underwent both CT and MR imaging procedures (n=96). The screening process led to the exclusion of several patients. Specifically, patients with osteoarthritis secondary to fracture (n=7), rheumatoid arthritis (n=3), osteonecrosis (n=1), as well as those who had undergone knee surgeries (n=3) were removed from the pool. Additionally, patients whose scans were compromised due to poor image quality were also excluded (n=2). Following this screening process, our final study cohort comprised of 80 patients. The detailed selection process is visualized in Figure 1. All participants completed the standardized Western Ontario and McMaster Universities Arthritis Index questionnaire for pain, stiffness, and functional impairment to evaluate the severity of knee symptoms. According to the Kellgren-Lawrence grade (KL grade) and clinical symptoms, subjects were divided into three groups: normal control (n=30, KL=0), mild OA (n=25, KL=1-2), and advanced OA (n=25, KL=3-4). The acquisition conditions of CT and MR are provided in supplementary details. Demography data are presented in Table 1.

### Study Designs

**Image Preprocessing**

CT bed and clothing were removed through threshold segmentation. The MR images were cropped to a size of 640×640×120 and CT images were cropped to 150×150×90 as Region of Interest (ROI). For the training of the distillation-learning-based regression module, CT image patches are randomly selected from trabecular areas with a size of 12×12×12, and the corresponding MR image patches are extracted and resized to 48×48×16.

**SRRD Framework**

The framework of SRRD for cross-modal trabecular parameter regression is presented in Figure 2, which consists of two major modules: (1) the synthesis and registration module for cross-modal registration, and (2) the distillation learning guided regression module for analyzing trabecular parameters at the patch level. The synthesis and registration module intends to establish anatomical patch-to-patch correspondences between CT and MR. Then, the paired MR and CT patches are fed as inputs for the distillation-learning-based regression module. The regression module is trained in two stages to enhance the CT-based student model's capability to regress trabecular parameters.

**Synthesis and Registration module**

To address the misalignment between the MR and CT images from the same patient, we designed the synthesis and registration module composed of three sub-modules, which sequentially address the misalignment in FOV and resolution, intensity distribution, and knee-bent angles. Below are the details of three sub-modules:

*Coarse registration sub-module.* — First, a coarse registration sub-module is employed to overcome potential alignment issues caused by differences in FOV and resolution during MR-to-CT synthesis. The sub-module utilizes ANTs(19) to generate coarsely-registered MR (crMR) images with the same resolution and close FOV as CT images.

*Synthesis sub-module.* — The goal of this sub-module is to generate synthesized CT images (sCT) with similar intensity distribution to CT from crMR images while maintaining its anatomical structure. Referred from Generative Adversarial Network (GAN)(20), we design a Structural-Preserving Synthesis Network (SPSNet) for synthesis task by the following strategies: (1) we use a pretrained HED(21) network to extract edges as outlines for bones, and the edge information is concatenated with real or synthesized images, (2) we adopt mutual information (MI)(22) loss between synthesized images and real images to increase consistency in edges and structure. The MI loss is calculated as follows:

$$L_{MI} = \sum_{y_1 \in I_{sCT}} \sum_{x_1 \in I_{crMR}} p(x_1, y_1) \log \frac{p(x_1, y_1)}{p(x_1)p(y_1)} + \sum_{y_2 \in I_{sMR}} \sum_{x_2 \in I_{CT}} p(x_2, y_2) \log \frac{p(x_2, y_2)}{p(x_2)p(y_2)}$$

Here, $I(\cdot)$ denotes images of the correspondent modality, $p(\cdot)$ denotes the intensity distribution of the image, and $p(\cdot, \cdot)$ denotes the joint intensity distribution of the two images.

*Refined registration sub-module.* — The refined registration sub-module addresses the difference in knee-bent angles by first segmenting femur and tibia, which were then registered separately using rigid transformation by ANTs(19) software. This produces a warping field for each component. Finally, the accurate warping field between MR and CT is derived through the three sub-modules described above, and the refined-registered MR images (rrMR) are produced.

**Distillation-learning-based Regression module**

Upon achieving accurate registration between two modalities, it becomes feasible to access paired MR and CT patches aligned for the same anatomical regions as necessities for distillation learning. Trabecular parameters, calculated from MR patches and serving as ground-truth, are expected to be regressed from the correspondent CT patches. To overcome the limitations of low-resolution CT modality, our proposed regression module leverages the high-resolution MR patches as a guide for training the CT-based model through distillation learning.

The distillation-learning-based training process involves two stages: (1) the pre-training stage, where the teacher model is trained to regress the trabecular parameters of MR patches accessed from morphology calculation; and (2) the distillation stage, where the teacher model guides the CT-based student model through knowledge distillation. During inference, only the student model is used for regression, as MR images are unavailable. Note that in the distillation stage, the distillation loss ($L_d$) is computed both on regressed parameters ($p$) and feature maps ($M$) as follows:

$$L_d = w_1 ||p_s - p_t|| + \sum_{i=1}^{k} w_{2,i} |M_{s,i} - M_{t,i}|,$$

Here, $s$ denotes the student model, $t$ denotes the teacher model, and $i$ denotes the $i$-th feature map. The $L_2$ loss on predictions regularizes the student by bringing the its prediction closer to that of the teacher. Additionally, the $L_1$ loss on intermediate feature maps further enhances the similarity between features extracted from the student and teacher.

**KOA classification based on regressed parameters**

Upon completing the regression of patch-level trabecular parameters, our objective is to access patient-level KOA classification for diagnostic purposes. Initially, we partitioned each trabecular region of the femur or tibia into six distinct subregions. Subsequently, we calculated the mean and standard deviation of the regressed trabecular parameters among patches within each subregion, which are utilized as input features for KOA classification using Supported Vector Machine (SVM)(23) as referred from (24).

*Implementation*

There are four networks trained in this framework: (1) SPSNet for MR-to-CT synthesis, (2)HED network(21) for edge-attraction, (3)2.5D U-net(25) for segmentation; (4) Distillation-learning-based regression network using Resnet50(26) as backbones. The networks are implemented in Pytorch(27) and trained on a single NVIDIA Tesla A100 GPU. More training details are provided in the supplementary materials. Code is available at https://github.com/jackhu-bme/SRRD.

*Statistical analysis*

Statistical analyses were performed using Scipy(28) package in Python, and results are reported with mean and standard deviation, derived from a five-fold cross-validation process. To evaluate the accuracy of registration, landmarks in MR and CT are identified and the target registration errors (TRE) over them are computed. To assess the regression performance, the mean intraclass correlation coefficients (ICCs) of trabecular parameters were reported. A lower TRE (higher ICC) indicates better performance for registration(regression). The comparisons of methods on registration and regression were conducted using paired t-test. The trabecular parameters studied consist of Bone Volume/Total Volume (BV/TV), Trabecular Thickness (Tb.Th), Trabecular Separation (Tb.Sp), and Trabecular Number (Tb.N). When evaluating the KOA classification performance, the metrics include the precision, recall, F1 score and AUCs with its 95% confidence intervals (CI). Statistical significance was assessed using a 5% error threshold.

# Results

To evaluate the effectiveness of our proposed models, we sequentially compared the performances of different methods on MR-to-CT synthesis, cross-modal registration, trabecular parameter regression, and KOA classification.

*Registration results*

**MR-to-CT synthesis**

As the synthesis sub-module aims to perform modality transformation while retaining anatomical structures, we commence by comparing the anatomical similarity between sCT and crMR, and image intensity distribution consistency between sCT and CT. The qualitative synthesis results are shown in Figure 3, which compares our SPSNet with CycleGAN(29) and MUNIT(30). It can be observed that CycleGAN fails to maintain anatomical structures of crMR, and distinct distortion exist in the femur in sCT by both MUNIT and CycleGAN. Nevertheless, our proposed synthesis module preserves more anatomical structure and ensures successful modality transformation from crMR.

To further quantify the anatomical structural similarity between sCT and crMR, the TRE of landmarks is tabulated in Table 2. Significantly lower TRE (2.93±1.75mm) is observed on sCT images generated by our proposed SPSNet than CycleGAN (6.70 ± 4.22mm) and MUNIT (4.03 ± 2.81mm).

**Overall MR-to-CT registration**

With the incorporation of synthesis technique, we intend to implement the overall MR-to-CT registration which is to align the two modalities that have differences in the resolution, intensity distribution, and knee-bent angles.

Qualitative comparisons of the overall registration are shown in Figure 4 While the tibia region is well registered in crMR, unsatisfied registration performance is observed in the femur due to the differences of angles between the femur and tibia. Our proposed refined registration sub-module lets the femur and tibia register separately, resulting in more aligned locations in the refined-registered MR (rrMR).

Table 2 also presents the ablation study of the proposed registration module using TREs. Compared to the coarse registration by ANTs(19), SPSNet performs modality transformation and reduces the misalignment caused by intensity distribution differences in registration, with a lower TRE of 8.06±4.98mm (vs. 13.56±8.41mm, $P < .001$). Moreover, the refined registration sub-module aligns knee-bent angles, resulting in a significantly lower TRE of 3.67±2.43mm ($P < .001$).

*Distillation-learning-based Regression*

The aim of the regression module is to regress trabecular parameters on CT patches, and ICCs between regressed parameters and ground-truth are listed in Table 3. The ICCs of different regression models were compared, and the CNN-based models achieved higher ICCs (0.545 ± 0.067) than the radiomic-feature-based model (0.409 ± 0.024). With the distillation learning technique, CT-based regression model can further achieve ICCs of 0.742 ± 0.046 when guided by a well pre-trained MR-based teacher model, as shown in Figure 5.

*KOA Classification Based on Regressed Trabecular Parameters*

We further conducted comparisons of KOA classification based on regressed trabecular parameters between alternative methods. After partitioning and utilizing SVM (23), and the statistical results of classification are presented in Table 4. The proposed SSRD method show significant better performance than the CNN-based method ($P < .001$).

# Discussions

In this paper, we proposed the distillation-learning-based method named SRRD for trabecular parameter regression and investigated the feasibility of CT-based subchondral bone microstructural analysis. The proposed SRRD method encompasses a distillation-learning-based regression module in conjunction with a synthesis and registration module. The design of the regression module aims to leverage MR to guide trabecular parameter regression on CT via distillation learning. To provide correspondent MR and CT patches as basis of distillation learning, the registration module is proposed to register the two modalities. Experimental results show that incorporation of distillation learning remarkably improves the trabecular parameter regression performance of CNNs in SRRD. Moreover, SRRD significantly elevates KOA classification performance based on regressed parameters, demonstrating its utility in CT-based OA diagnosis.

In our study, accurate MR-CT registration yields anatomically-correspondent MR and CT patch pairs, which serves as a prerequisite for distillation learning. While previous methods of knee MR-CT registration relies heavily on segmentation labels(31, 32), our SRRD performs unsupervised registration based on inherent anatomical structures. The proposed synthesis and registration module in SRRD overcomes modality differences step-by-step. The coarse-registration submodule aligns resolution and FOV, the SPSNet sub-module performs cross-modality synthesis, and the refined-registration submodule aligns knee-bent angles. Note that our SPSNet maintains the most anatomical similarity relative to CycleGAN(29) and MUNIT(30) as shown in Table 3. This facilitates the preservation of physical structure locations during the synthesis, which is encouraged by the application of mutual information loss function. Furthermore, by extracting edge information, the structural consistency is further constrained. The SPSNet simplifies MR-CT registration task into registration between sCT and CT. In this way, the refined registration sub-module warped the femur and tibia separately to address the knee-bent angel inconsistency, resulting in a lower TRE of 3.67±2.43mm vs. 8.06±4.98mm ($P < .001$).

Based on the well-aligned CT and MR patches, the distillation learning technique is designed to enhance the CT-based regression task. Although prior approaches to subchondral bone microstructural analysis have utilized radiomic-based methods(24) and CNN-based methods(33), their direct application to CT yields unsatisfied results, as they focus on mono-modal predictions(24, 33). As the ground-truth comes from MR patches, the relatively-low resolution of conventional CT and intensity distribution differences between modalities contribute to the difficulty of regression. In this way, our main contribution is to assist CT-based model with distillation learning technique. Based on the demonstrated capacity of distillation learning to augment medical image segmentation on specific modalities(18), we incorporated it into cross-modal parameter regression. The distillation learning has significantly optimized CT-based CNN models ($P < .001$), which showed the effectiveness of harnessing rich information contained in high resolution MR patches. Despite the substantial modality and resolution differences between MR and CT, mean ICCs of CT-based regression by SRRD reaches $0.742 \pm 0.046$, illustrating the feasibility of CT-based subchondral bone microstructural analysis.

Additionally, the potential for KOA diagnosis based on regressed parameters is investigated. We refer from (24) and employed SVM(23) to classify KOA using regressed parameters. Among CT-based methods for KOA classification, the proposed SSRD yields significantly higher AUC than radiomic-based method ($P < .001$) and vanilla CNN-based method ($P < .001$). This enhanced performance is attributed to the superior regression performance of SRRD, which can provide more effective patch-level trabecular features for classification. However, it's noteworthy that gap exists between our CT-based SRRD method and MR-based method ($P = .003$) due to the inevitable information loss resulting from significant resolution discrepancies between MR and CT. Nonetheless, our proposed SSRD reaches the AUC score of 0.767 (95% CI, 0.681-0.853), which still demonstrates the significant potential for CT-based KOA diagnosis in the clinical scenario, given the low-cost acquisition of CT compared with MR.

The potential limitations and remedies in our studies are listed as follows: Firstly, the sample size is small and from a single center. Larger datasets from multiple centers are further needed to validate the effectiveness of our proposed method. Secondly, the classification performance by SVM (23) is also limited, and replacing SVM (23) with deep neural networks in future research on larger datasets could further improve the classification accuracy. Additionally, current OA classification is based on the overall regressed parameters of 6 distinct regions, which is implemented in a conventional manner. We will continue to explore the feasibility of employing all the regressed patch-level parameters comprehensively for classification, which is expected to further improve the performance based on more comprehensive features. Finally, while the distillation module transfers feature extraction knowledge from the teacher to student through feature map losses, other techniques such as contrastive learning (34) can be explored to further enhance the regression process.

In conclusion, we showed that our distillation-learning-based SRRD method reliably regresses trabecular parameters on CT images, therefore significantly improves CT-based KOA classification. Distillation-learning-based algorithm, SRRD, tackles challenges for trabecular parameter regression on cost-effective yet low resolution CT images. This capacity paves the way for the expansion of subchondral bone microstructural analysis in clinical applications, thereby fostering advancements in the diagnosis and assessment of KOA.

# Tables

**Table 1**: Demography analysis of patients with various stages of OA.

| Variables   | Total (n=80)   | Normal (n=30) | Mild OA (n=25) | Advanced OA(n=25) | P value |
|---|---|---|---|---|---|
| Ages (Years) | 51.3 ± 14.5 | 38.6 ± 9.6 | 56.3 ± 9.7 | 62.4 ± 10.1 | < 0.001 |
| Sex |  |  |  |  | 0.58 |
|   Male | 45(56.2%) | 15(33.3%) | 14(31.1%) | 16(35.6%) |  |
|   Female | 35(43.8%) | 15(42.9%) | 11(31.4%) | 9(25.7%) |  |
| BMI | 25.2 ± 3.1 | 24.6 ± 2.7 | 24.9 ± 3.6 | 26.1 ± 2.6 | 0.16 |
| Knee |  |  |  |  | 0.78 |
|   Left knee | 41(51.2%) | 14(34.1%) | 13(31.7%) | 14(34.1%) |  |
|   Right knee | 39(48.8%) | 16(41.0%) | 12(30.8%) | 11(28.2%) |  |

**Note.** —BMI = body mass index.

Ages and BMI are cited as mean values ± standard deviations.

**Table 2**: The TRE comparison for the MR-to-CT synthesis and the overall registration.

| | MR-to-CT Synthesis (crMR-sCT) | | |
|---|---|---|---|
| Methods | CycleGAN(29) | MUNIT(30) | **SPSNet(ours)** |
| TRE/mm | 6.70±4.22 | 4.03±2.81 | **2.93±1.75** |
| | Overall Registration (MR-CT) | | |
| Methods | ANTs (rigid only) (19) | ANTs (rigid only) (19) + SPSNet | **ANTs (rigid only) + SPSNet + refined registration (ours)** |
| TRE/mm | 13.56±8.41 | 8.06±4.98 | **3.67±2.43** |

**Note.** —In the synthesis process, a lower TRE indicates a smaller spatial gap between the same landmark in the synthesized and real images. For the overall registration, lower TREs indicates better registration results.

**Table 3**: ICC comparison of trabecular parameter regression results.

|         | Radiomic-feature-based model | CNN regression model | Distillation-learning-based CNN regression model |
|---------|------------------------------|----------------------|--------------------------------------------------|
| BV/TV   | 0.548 ± 0.011                | 0.670 ± 0.053        | 0.804 ± 0.037                                    |
| Tb.Th   | 0.407 ± 0.032                | 0.488 ± 0.038        | 0.773 ± 0.042                                    |
| Tb.Sp   | 0.356 ± 0.017                | 0.523 ± 0.054        | 0.711 ± 0.063                                    |
| Tb.N    | 0.326 ± 0.036                | 0.502 ± 0.085        | 0.622 ± 0.133                                    |
| Overall | 0.409 ± 0.026                | 0.545 ± 0.059        | 0.742 ± 0.078                                    |

**Note.** —The compared regression methods include the radiomic-feature-based model, CNN regression model, and the proposed distillation-learning-based CNN regression model. The trabecular parameters that were regressed are BV/TV, Tb.Th, Tb.Sp, and Tb.N. BV/TV = bone volume/total volume, Tb.Th = trabecular thickness, Tb.Sp = trabecular separation, Tb.N = trabecular number.

Table 4: The statistical analysis of trabecular-parameter-based KOA classification

|  | Precision | Recall | F1 score | AUC |
|---|---|---|---|---|
| Radiomic-based (CT) (24) | 0.579 ± 0.090 | 0.703±0.034 | 0.635±0.057 | 0.622 (0.550-0.694) |
| CNN (CT) (33) | 0.625 ± 0.043 | 0.684±0.056 | 0.652±0.039 | 0.658 (0.574-0.742) |
| **SRRD (CT)** | **0.743 ± 0.072** | **0.793 ± 0.062** | **0.771 ± 0.056** | **0.767 (0.681-0.853)** |
| CNN (MR) | 0.786 ± 0.061 | 0.865 ± 0.037 | 0.824 ± 0.044 | 0.812 (0.756-0.868) |

**Note.** —The classification is implemented using Supported Vector Machines (SVM) (23) on MR and CT image patches. The classification comparisons of different CT-based methods are shown in the table, and the MR-based method using CNN as the backbone is provided in the last row for reference. The precision, recall, F1 score and AUC are calculated for each method, and 95% confidence intervals of AUCs are provided. AUC = area under the receiver operating characteristic curve, CI = confidence interval.

**Table S1:** Comparison of FOV, resolution and scanner of MR and CT images

|  | MR image | CT image |
|---|---|---|
| In-plane FOV | 640 × 640 | 150 × 150 |
| Resolution (mm$^3$) | 0.234 × 0.234 × 1.500 | 0.977 × 0.977 × 1.000 |
| Scanner | Philips Healthcare Achieva 3.0TX | Philips Brilliance 64 CT |

**Table S2:** The training details of networks in our proposed method

| Networks | Dimension | Learning Rate | Optimizer | Epochs | Dataset | Data Augmentation |
|---|---|---|---|---|---|---|
| SPSNet | 2.5D | 0.0002 | Adam(35) | 100 | crMR, CT | Shift, Flip, Zoom |
| HEDnet(21) | 2D | 0.0001 | Adam(35) | 30 | BSDS500 | Shift, Flip |
| Unet(25) | 2.5D | 0.0001 | Adam(35) | 50 | CT | Shift, Flip |
| Distillation-learning-based regression module | 3D | 0.0003 | Adam(35) | 50 | MR, CT | Flip |

# Figure Legends

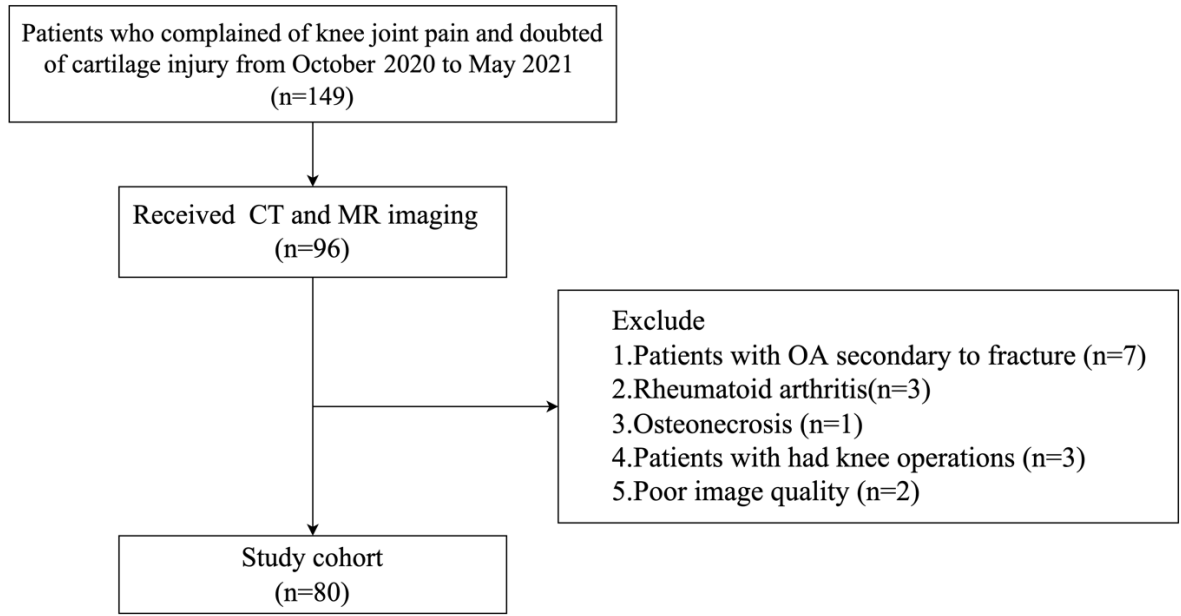

**Figure 1: Flowchart illustrating the patient selection process in this study.**

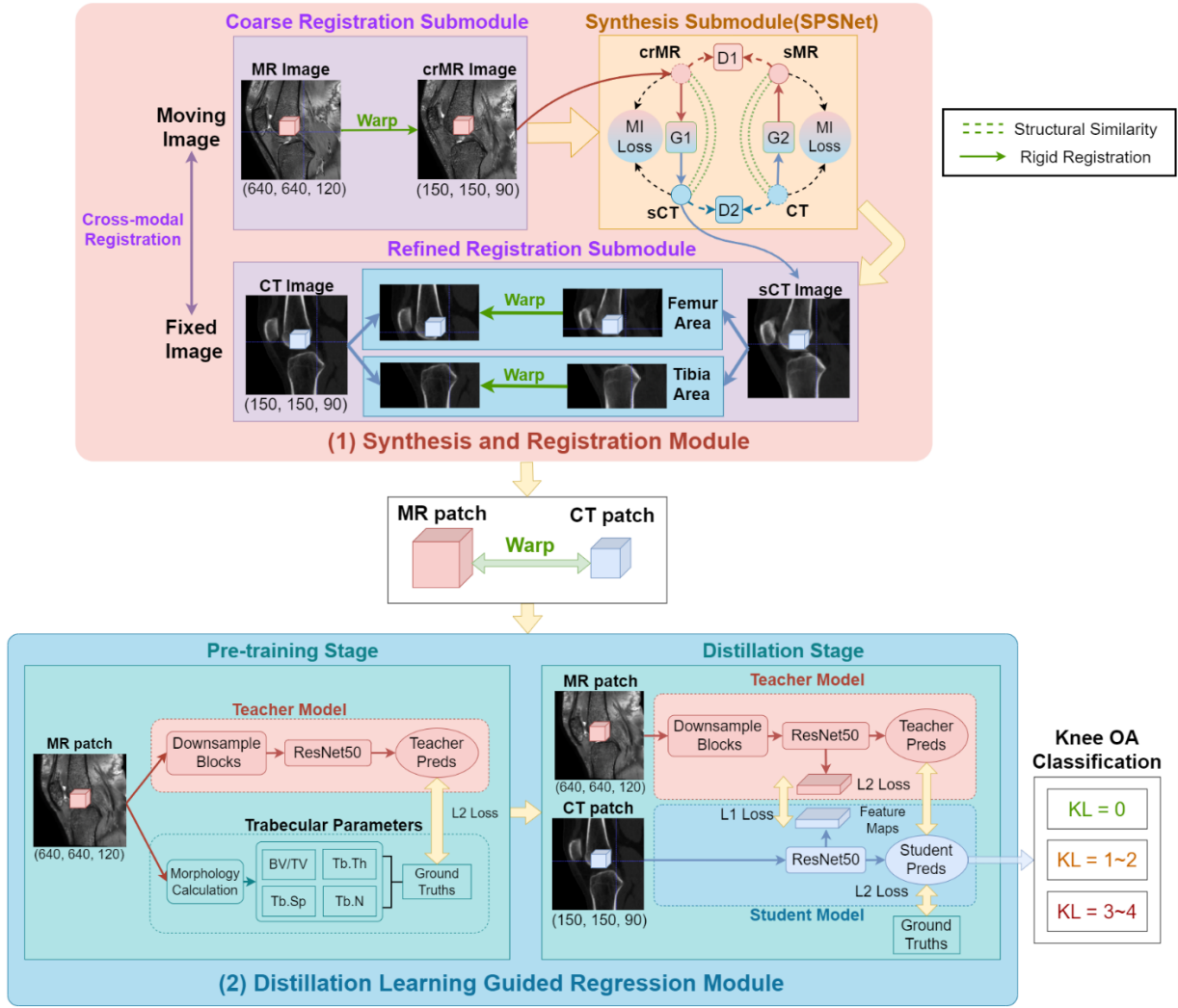

**Figure 2:** Framework of SRRD for cross-modal trabecular parameter regression and subchondral bone microstructural analysis. The synthesis and registration module performs MR-to-CT registration in three steps: 1) coarse MR-to-CT registration to generate crMR images, 2) modality transformation through SPSNet on crMR images to synthesize sCT with anatomical structure preserved, 3) refined registration from sCT to CT on each bone opponent. With paired MR and CT patches, the distillation-learning-based regression module first pre-trains an MR-based teacher model and then supervises the CT-based student model with knowledge distillation. The regressed parameters are further employed for KOA classification. crMR = coarsely-registered MR, sCT = synthesized CT.

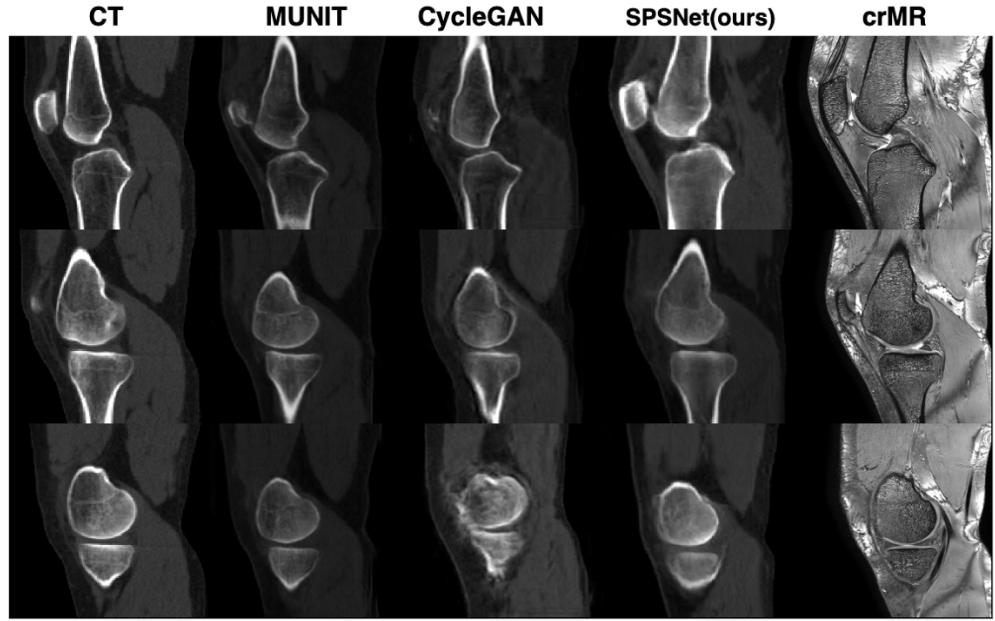

**Figure 3**: Comparison of crMR-to-sCT synthesis of different methods. Each row displays the synthesis results of a sampled slice from the test dataset. The columns, arranged from left to right, represent CT images, sCT images by MUNIT(30), sCT images by CycleGAN(29), sCT images by our proposed SPSNet and the original crMR images, respectively. crMR = coarsely-registered MR, sCT = synthesized CT.

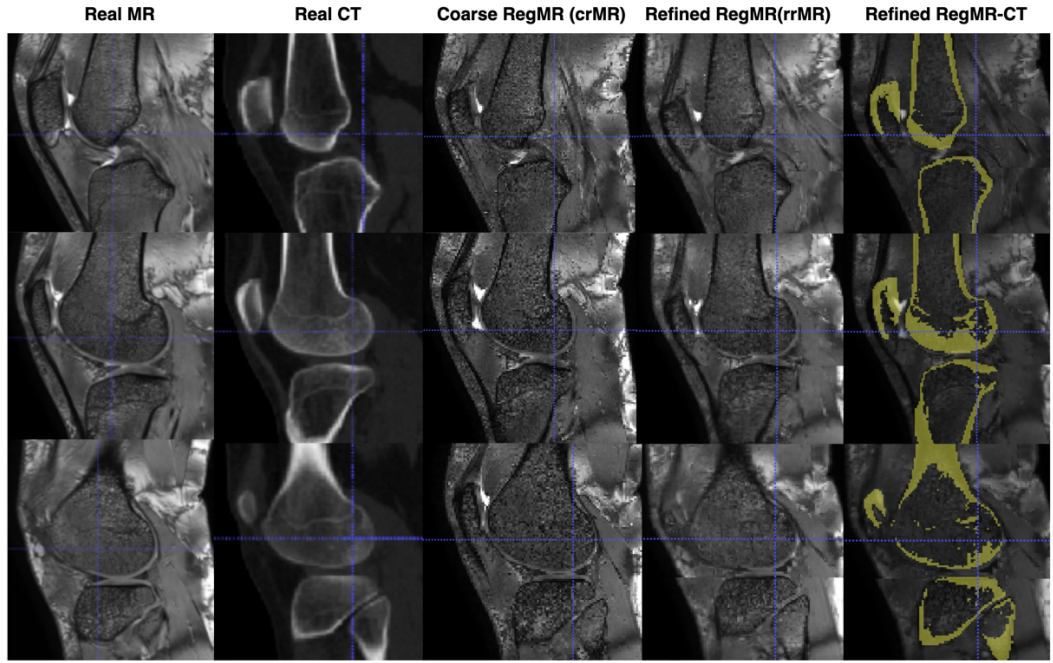

**Figure 4:** Comparisons of different registration methods. Each row shows the synthesis results of a sampled slice from the test dataset. From left to right, the columns are sequentially real MR images, real CT images, coarsely-registered MR (crMR) images by ANTs (rigid only) (19), refined-registered MR images (rrMR), and the visualization of registration performance between rrMR and real CT. During the generation of the visualization images, the high signals on the registered CT images are colored yellow and overlaid onto the corresponding MR images. The visualization shows a satisfied structural alignment situation of rrMR and CT images.

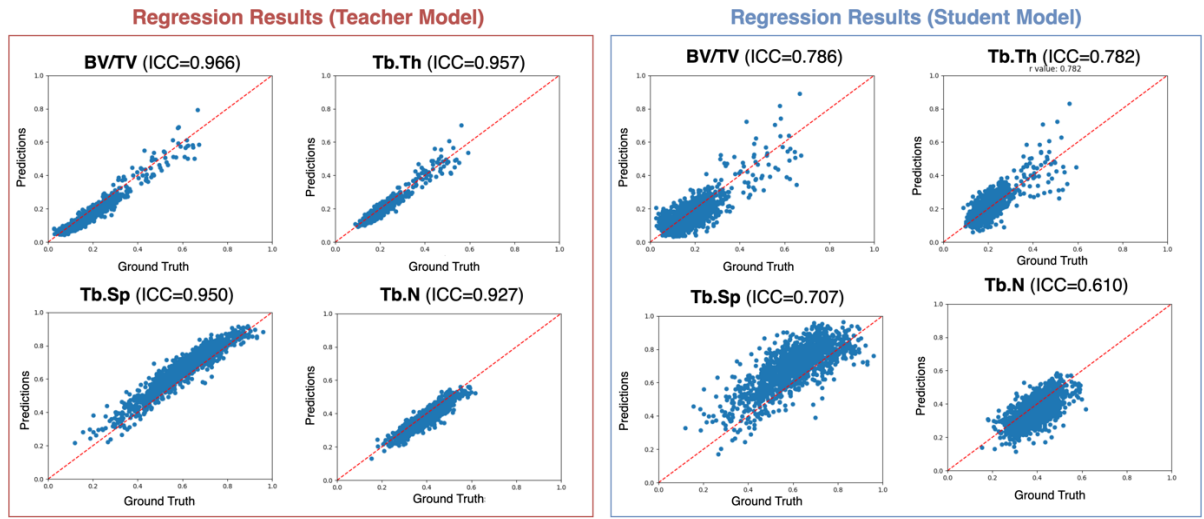

**Figure 5**: Scatter plots of trabecular parameter regression results. Both the regression results from the teacher (left) and student (right) models with four plots representing BV/TV, Tb.Th, Tb.Sp, and Tb.N are presented. Each blue point represents a regression result from a single CT patch in the trabecular section of the knee of one patient. BV/TV = bone volume/total volume, Tb.Th = trabecular thickness, Tb.Sp = trabecular separation, Tb.N = trabecular number.

# Supplementary details

## 1 Acquisition conditions for CT and MR

### 1.1 CT acquisition

CT acquisition and sagittal reformation of the knee were acquired using 64-slice spiral multi-detector CT (Brilliace64, Philips). The scan protocol was as follows: 120kV, 200mAs, slice thickness 1mm, reconstruction increment 0.3mm, field of view [FOV] 14cm, scan length of 30cm. The CT data were reconstructed with the U70u kernel. Central quality control of all CT examinations was performed by the same radiologist. (Q.G.)

### 1.2 MR acquisition

All participants were scanned using a 3T MRI scanner (Philips Healthcare Achieva 3.0TX) with an eight-channel knee coil (Philips Healthcare). The knee flexion angle was adjusted to 20-30°, and an immobilization sponge was used to increase participant comfort and reduce motion artifacts. The MRI-based trabecular subchondral bone protocol was as follows: A sagittal 3D balanced fast-field echo (3D BFFE) sequence (repetition time/time to echo [TR/TE] = 10/5.0, field of view [FOV] = 14 cm, matrix = 640 × 640, flip angle =15°, in-plane spatial resolution = 0.234 mm × 0.234 mm, slice thickness= 1.5 mm, sensitivity encoding [SENCE] = 2, scan time = 8 min 14 s).

### 1.3 Comparisons of pre-processed MR and CT

The comparisons of pre-processed MR and CT in in-plane FOV, resolution and scanner are listed in the Table S1.

## 2 Implementation details of networks

### 2.1 Training details of networks in SRRD

The training details of networks in different modules of SRRD are listed in the Table S2. The details include the dimension, learning rate, optimizer, training epoch, dataset, and the applied data augmentations.